\documentclass[twocolumn,twoside,slac_two]{revtex4}
\usepackage{longtable}
\usepackage{graphicx}
\usepackage{fancyhdr}
\usepackage[]{natbib}
\begin{document}

\title{Photopolarimetric monitoring of 41 blazars in the optical and
 near-infrared bands with the Kanata telescope}

%

\author{Y. Ikejiri, M. Uemura, M. Sasada, K. Sakimoto, R. Ito,
M. Yamanaka, A. Arai, Y. Fukazawa, T. Ohsugi, K.S. Kawabata}
\affiliation{Hiroshima University}
\author{S. Sato, M. Kino}
\affiliation{Nagoya University}

\begin{abstract}
Blazars are a kind of active galactic nuclei (AGN) in which a
 relativistic jet is considered to be directed along the line of sight.
They are characterized by strong and rapid variability of the flux and
 high polarization.
We performed a monitoring of 41 blazars in the optical and near-infrared
 regions from 2008 to 2009 using TRISPEC attached to the ``Kanata''
 1.5-m telescope.
In this paper, we report the correlation of the flux, color
 and polarization using our data, and discuss universal features for
 blazars, which have not fully been established.
Three blazars (3C~454.3, QSO~0454$-$234, and PKS~1510$-$089) tended to be
 redder when they were brighter, only during their faint states.
This color behavior suggests that the contribution of a thermal
 component is strong in the faint states for those objects.
Excluding this ``redder-when-brighter'' phase, we found that 24 blazars
 tended to be bluer when they were brighter. 
This number corresponds to 83$\%$ among well-observed objects which we
 observed for $>10$ nights.
Thus, we conclude that the ``bluer-when-brighter'' trend is a universal
 feature for blazars.
On the other hand, the correlation of the flux and the polarization
 degree is relatively weak; only 10 objects showed a significant positive
 correlation.
We also investigated the luminosity-dependence of the color and
 polarization, and found that lower luminosity objects have
 smaller variation amplitudes both in the flux, color, and polarization
 degree.

\end{abstract}

\maketitle

\thispagestyle{fancy}


\section{Introduction}
Blazars are one of subgroups of active galactic nuclei (AGN).
They have a relativistic jet which is considered to be directed along
 the line of sight.
The emission from blazars can be observed in a very wide range of
 wavelength from the radio to TeV $\gamma$-ray regions, with a strong
 variability in various timescales.
Their spectral energy distribution (SED) is characterized by non-thermal
 continuum spectra with broad low and high energy components.
The low energy component is believed to be synchrotron radiation by
 relativistic electrons in the jet.
The origin of the high energy component is not fully understood, while
 the most plausible source is the emission via inverse Compton
 scattering of the synchrotron emission and/or external photons.

Blazars have been classified into three categories: high-energy peaked
 BL Lac object (HBL), intermediate-energy peaked BL Lac object (IBL) and
 low-energy peaked BL Lac object (LBL). 
The SED of the synchrotron component has a peak at a frequency lower
 than the optical region in LBL, while it is higher in HBL.
The peak frequency is around the optical---infrared region in IBL.

The mechanism of variations in blazars has been extensively studied with
 the variation in color.
\cite{Carini1992} reported a possible feature
 that both BL~Lac (LBL) and OJ~287 (IBL) became bluer when they were brighter
 (\cite{Clements2001}).
\cite{Ghisellini1997} observed S5~0716$+$714 (IBL), and found that a
 bluer-when-brighter feature was seen only in its ``low'' state.
In this object, the bluer-when-brighter feature was later found in
 variations having a time-scale shorter than a few days in its ``high''
 state (\cite{Wu2007}; \cite{Sasada2008}).
\cite{Villata2002} and \cite{Villata2004} reported that the
 bluer-when-brighter feature in BL~Lac was only observed for short-term
 variations, and not prominent in long-term ones.
\cite{Raiteri2001} performed multi-band photometric observations of
 AO~0235$+$164 (LBL) for four years, and found a bluer-when-brighter trend in
 this object.

Systematic studies for several blazars have also been performed in order
 to establish a common behavior of color variations in blazars.
\cite{Ghosh2000} performed the observations of five blazars.
3C~66A (IBL) only exhibited a bluer-when-brighter trend in their sample,
 while the other objects showed no significant correlation of the flux
 and color.
\cite{Gu2006} investigated the color variation of eight blazars.
Bluer-when-brighter features were confirmed in five blazars.
Among the other three objects, 3C~345 (LBL) showed no correlation of the
 flux and the color was detected, and 3C~454.3 (FSRQ) and PKS~0420$-$01 (FSRQ)
 showed a reddening trend when they were brightening.
This trend is called ``redder-when-brighter'', and has been confirmed in
 3C~454.3 (\cite{Raiteri2008}, \cite{Villata2006}).
Thus, it is currently unclear whether the bluer-when-brighter feature is
 a universal one in blazars.

Only few cases have been reported in which the polarization degree
 correlates with the flux.
In general, the temporal variation of polarization is considered to be
 erratic in blazars (e.g. \cite{Moore1982}).
The polarization degree of Mrk~421 (HBL) increased to $\sim 14$~\% associated
 with its outburst in 1997 (\cite{Tosti1998}).
A significant correlation of the flux and the polarization degree
 was also detected in AO~0235$+$164 in 2006 (\cite{Hagen2008}).
Dense and long-term polarimetric observations are required in order to
 find a possible universal relation of the flux and polarization in
 blazars.

The {\it Fermi} Gamma-ray Space Telescope was launched in June, 2008.
{\it Fermi} can measure a $\gamma$-ray spectrum from 30~MeV up to 300GeV.
At this occasion, we performed the photopolarimetric monitoring of
 blazars in the optical and near-infrared bands using TRISPEC attached
 to the Kanata telescope of the Higashi-Hiroshima Observatory. 
In this paper, we report the results of the observations with Kanata.
We investigated the relation of the luminosity, color and polarization
 degree.

\section{Observation}

\begin{table}[]
\begin{center}
\caption{Observation log.\\
The columns are :(1)the object name; (2)the blazar class; (3)the observation period;
 (4)the number of observation nights.} 
\begin{tabular}{llll}
\hline
\textbf{object} & \textbf{class} & \textbf{observation period} &
\textbf{n} \\
\textbf{(1)} & \textbf{(2)} & \textbf{(3)} & \textbf{(4)} \\
\hline
Mis V1436 & FSRQ & 08 Dec.17 - 09 Nov.28 & 90 \\
PKS 0215+015 & FSRQ & 08 Sep.28 - 09 Aug.25 & 6 \\
QSO 0324+341 & FSRQ & 08 Nov.25 - 09 Oct.14 & 2 \\
QSO 0454-234 & FSRQ & 08 Oct.14 - 09 Nov.21 & 53 \\
OJ 49 & FSRQ & 08 Oct.21 - 09 Nov.29 & 38 \\
QSO 0948+002 & FSRQ & 09 Mar.29 - 09 Apr.10 & 3 \\
3C 273 & FSRQ & 08 Dec.10 - 09 Nov.29 & 59 \\
QSO 1239+044 & FSRQ & 09 Jan.29 - 09 Feb.16 & 7 \\
3C 279 & FSRQ & 08 Dec.16 - 09 Jul.14 & 61 \\
PKS 1502+106 & FSRQ & 08 Aug.14 - 09 Jun.26 & 71 \\
PKS 1510-089 & FSRQ & 09 Jan.26 - 09 Jul.22 & 52 \\
3C 454.3 & FSRQ & 08 May22 - 09 Nov.29 & 226 \\
1ES 0323+022 & HBL & 08 Jul.29 - 09 Oct.26 & 23 \\
1ES 0647+250 & HBL & 08 Sep.26 - 09 Mar.17 & 6 \\
1ES 0806+524 & HBL & 08 Oct.17 - 09 Nov.17 & 14 \\
Mrk 421 & HBL & 08 Jun.17 - 09 Mar.31 & 42 \\
ON 325 & HBL & 08 May31 - 09 Nov.23 & 45 \\
PG 1553+113 & HBL & 08 Jul.23 - 09 Sep.18 & 21 \\
H 1722+119 & HBL & 08 Jul.18 - 09 Oct.17 & 28 \\
1ES 1959+650 & HBL & 08 Jul.17 - 09 Nov.28 & 53 \\
PKS 2155-304 & HBL & 08 Jul.28 - 09 Nov.28 & 130 \\
1ES 2344+514 & HBL & 08 Jul.28 - 09 Oct.11 & 17 \\
PKS 0048-097 & IBL & 08 Oct.11 - 09 Sep.23 & 46 \\
S2 0109+224 & IBL & 08 Jul.23 - 09 Oct.29 & 74 \\
3C 66A & IBL & 08 Jul.29 - 09 Nov.29 & 194 \\
PKS 0422+004 & IBL & 08 Sep.29 - 09 Nov.11 & 42 \\
S5 0716+714 & IBL & 08 May11 - 09 Nov.29 & 203 \\
PKS 0754+100 & IBL & 08 Nov.29 - 09 Mar.18 & 28 \\
OJ 287 & IBL & 08 May26 - 09 Nov.29 & 191 \\
ON 231 & IBL & 08 Dec.29 - 09 Feb.02 & 5 \\
Mrk 501 & IBL & 08 May02 - 09 Sep.19 & 40 \\
PKS 1749+096 & IBL & 08 Jul.19 - 09 Sep.10 & 78 \\
3C 371 & IBL & 08 Jul.10 - 09 Nov.17 & 100 \\
AO 0235+164 & LBL & 08 Jul.17 - 09 Jul.22 & 71 \\
S5 1803+784 & LBL & 08 Jul.22 - 09 Oct.18 & 35 \\
BL Lac & LBL & 08 May18 - 09 Nov.27 & 189 \\
OQ 530 & unknown & 08 Jul.27 - 08 Sep.10 & 3 \\
PKS 1222+216 & unknown & 09 Apr.10 - 09 Apr.22 & 1 \\
RX J1542.8+612 & unknown & 09 May22 - 09 Nov.19 & 59 \\
S4 0954+65 & unknown & 08 Dec.19 - 09 Jan.07 & 2 \\
3EG 1052+571 & unknown & 08 Oct.07 - 09 Oct.13 & 3 \\
4C 14.23 & unknown & 09 Oct.13 - 09 Nov.26 & 14 \\
\hline
\end{tabular}
\label{tab:obslog}
\end{center}
\end{table}


We performed observations of 41 blazars in the optical $V$ and the
 near-infrared $J$ and $K_{\rm s}$ bands using TRISPEC attached to the
 1.5-m ``Kanata'' telescope at Higashi-Hiroshima Observatory since May
 2008.
We selected the monitored blazars from the catalog, ``Extended list of
206 possible AGN/blazar targets for GLAST multifrequency
analysis''\footnote{http://glastweb.pg.infn.it/blazar/}
 with an apparent magnitude of $R\lesssim 16$.
The number of the selected objects is 30 in this catalog.
New objects were occasionally added in our sample when {\it Fermi}
 detected a $gamma$-ray activity of them, or when optical flares were
 detected.
The number of the additional objects is 11.
Table~\ref{tab:obslog} lists our objects.

TRISPEC is capable of simultaneous three-band (optical one and
 near-infrared two bands) imaging or spectroscopy, with or without
 polarimetry (\cite{Watanabe2005}).
TRISPEC has a CCD and two InSb arrays.
An one polarization data point is obtained from four images which are
 obtained using the half-wave plate in the differential angles.
The magnitudes were measured using the aperture photometry technique.
Then, we calculate the differential magnitudes of objects with a
 comparison star taken in the same flame.
The magnitudes of the comparison stars were referred from \cite{Skiff2007},
 \cite{Hog2000}, \cite{Villata1998}, \cite{Gonzalez2001},
 \cite{SDSS}, \cite{Doroshenko2005}, \cite{Cutri2003}.

In a night, we obtained 12 sets of images for an object, from which three
 sets of polarimetric parameters were obtained. Exposure times depend on
 the sky condition and the apparent magnitude of objects, typically
 50--200~s, 5--20~s, and 2---5~s in the $V$, $J$ and  $K_{\rm s}$ band,
 respectively.
Since we make no discussion about intra-night variability in this paper,
 we only use nightly-averaged photometric and polarimetric data.

\section{Results}
 \subsection{Lightcurve and Color}

\begin{figure}[t]
 \includegraphics[width=130mm]{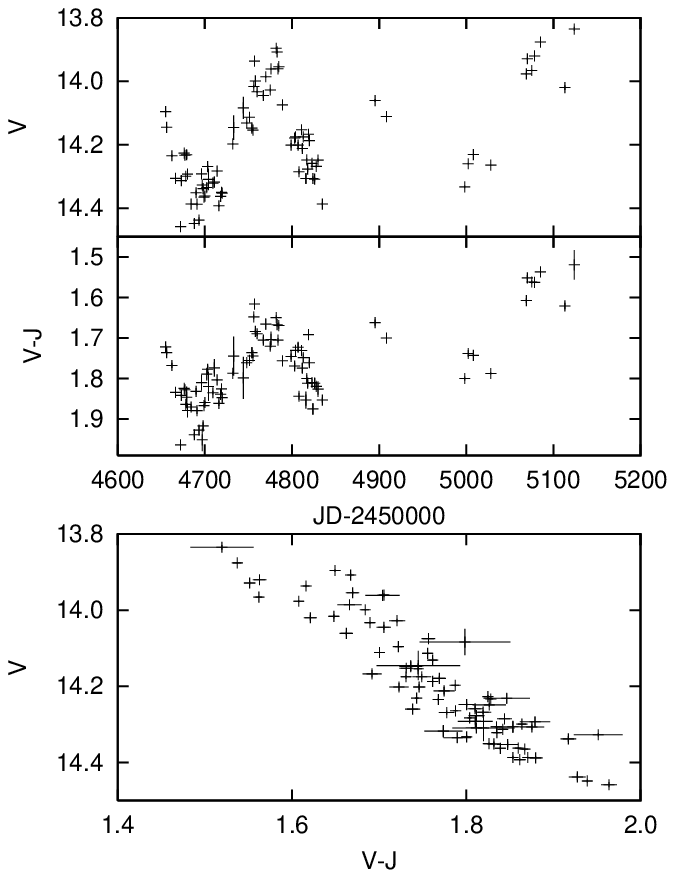}
 \caption{Our results for 3C~371.
 Top and middle panel: The lightcurve and the color
 variation. Bottom panel: The color-magnitude diagram.}\label{fig:3C371}
\end{figure}

\begin{figure}[t]
 \includegraphics[width=130mm]{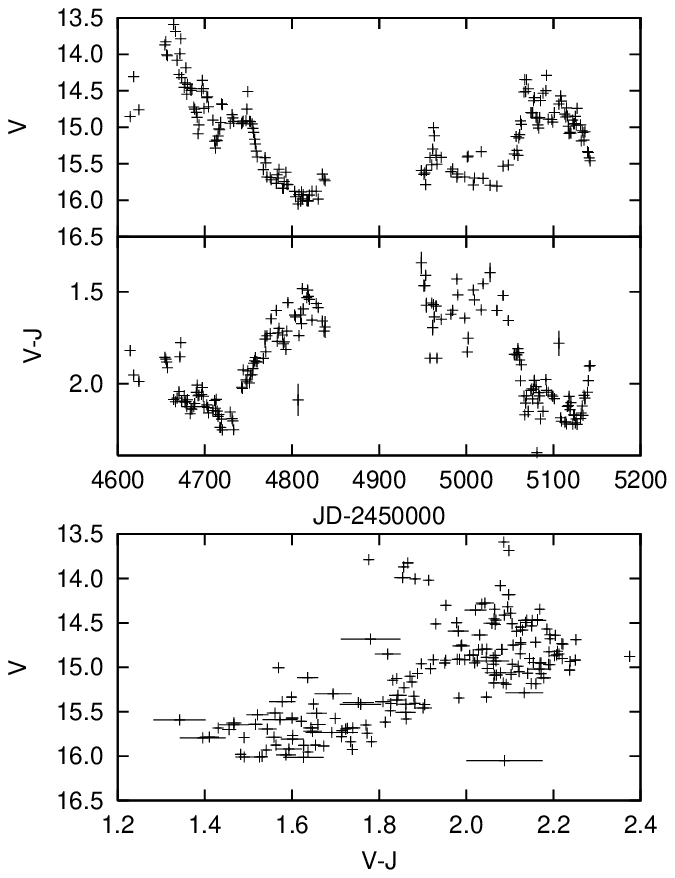}
 \caption{Our results for 3C~454.3.
 Top and middle panel: The lightcurve and the color variation. Bottom
 panel: The color-magnitude diagram.}\label{fig:3C454}
\end{figure}

\begin{figure}[t]
 \includegraphics[width=60mm]{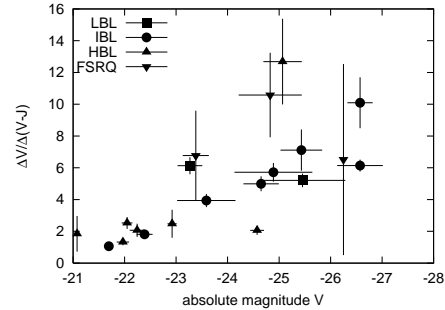}
 \caption{$\Delta V/\Delta (V-J)$ against the $V$-band absolute
 magnitude of 23 blazars. The absolute magnitudes are averages of all
 data we observed for each object. This figure only includes objects in
 which a significant bluer-when-brighter was
 detected.}\label{fig:abmag_v-vj} 
\end{figure}

\begin{figure}[t]
 \includegraphics[width=60mm]{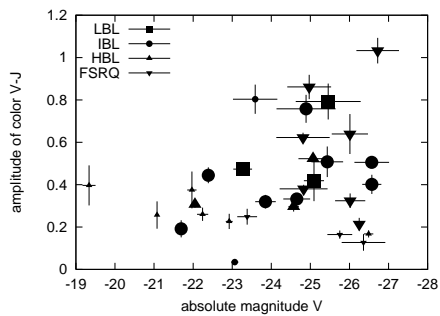}
 \caption{Peak-to-peak amplitudes of $V-J$ against the $V$-band absolute
 magnitudes of 38 blazars. The size of the symbol means the number of
 the observations. The small symbols: the objects observed for $<
 30$~d. The large symbols: the objects observed for
 $>30$~d.}\label{fig:abmag_col_peak2peak}
\end{figure}

We show an example of the light curve and the color-magnitude diagrams
 in figure~\ref{fig:3C371}.
This figure shows the result for 3C~371.
As shown in figure~\ref{fig:3C371}, the object became bluer when it was
 brighter. 
We calculated correlation coefficients between the magnitude and the
 color index for 41 blazars, and then performed $t$-test for the
 significance of the correlation. 
As a result, we found that 21 blazars exhibited a significant
 bluer-when-brighter feature.

We found that three blazars (3C~454.3, PKS~1510$-$089 and QSO~0454$-$234)
 tended to be redder when they were brighter for their faint states.
For example, figure~\ref{fig:3C454} shows the lightcurve and
 color-magnitude diagram of 3C~454.3.
A redder-when-brighter trend appeared when it was fainter than $V\sim
 15$.
Those three objects showed a significant bluer-when-brighter feature in
 the case that we only used the data brighter than $V=15$, $15.6$ and
 $16.0$ for 3C~454.3, PKS~1510$-$089 and QSO~0454$-$234, respectively.
The number of objects showing a bluer-when-brighter feature is, hence,
 24 with those three objects.

There are 17 blazars which showed no significant correlation
 between the $V$-band magnitude and the $V-J$ color.
The numbers of observation are quite small for most of these 17 objects;
 it is less than 10~d for 12 objects.
There are 5 blazars that was observed for $>10$~d and no significant
 correlation of the light curve and color: OJ~287, PG~1553$+$113,
 H~1722$+$119, S5~0716$+$784 and RX~J1542.8$+$612.
Without the objects which were observed for $<10$~d, the sample reduces
 to 29 blazars, among which 24 blazars showed significant
 bluer-when-brighter features.
This corresponds to 83~\% of the sample

Thus, our observation strongly indicates that the bluer-when-brighter
 feature is a common characteristics of most of blazar variations.

We calculated the slope in the color-magnitude diagram, 
 $\Delta V$/$\Delta (V-J)$, by fitting a linear function to the observed
 $V$ and $V-J$ of objects that exhibited a significant
 bluer-when-brighter feature.
$\Delta V$/$\Delta (V-J)$ is considered as an index of the flux
 variability against the color variation.
This can have an advantage to investigate the variability because the
 observed amplitudes of variations in $V$ and $V-J$ highly depend on the
 observation period.

Figure~\ref{fig:abmag_v-vj} shows $\Delta V$/$\Delta (V-J)$ against the
 average of the $V$-band absolute magnitude.
While there are 24 objects which showed a significant
 bluer-when-brighter feature, this figure includes 23 blazars because the
 redshift, z, of S2~0109$+$224 is unknown.
As can be seen in figure~\ref{fig:abmag_v-vj}, a lower luminosity
 object, like HBLs, tend to have a lower $\Delta V$/$\Delta (V-J)$.
This result suggests that the variation amplitude of the flux is smaller
 in lower luminosity objects.

Figure~\ref{fig:abmag_col_peak2peak} shows the relation of the observed
 amplitudes of the $V-J$ variation and absolute magnitudes in the $V$-band
 for 38 blazars whose $z$ is known.
In this figure, the size of the symbols represents the number of
 observations.
The objects observed for $<30$~d and $>30$~d are shown
 by the small and large symbols, respectively.
There is no objects which have both a low luminosity and a large
 amplitude.
This feature is still seen even if the objects that was observed for
 $<30$~d are excluded.
Hence, our observation suggests that the variation amplitudes are
 smaller in the objects with lower luminosities, like HBLs, both in the
 magnitude and the color.

 \subsection{Lightcurve and Polarization}

\begin{figure}[t]
 \includegraphics[width=130mm]{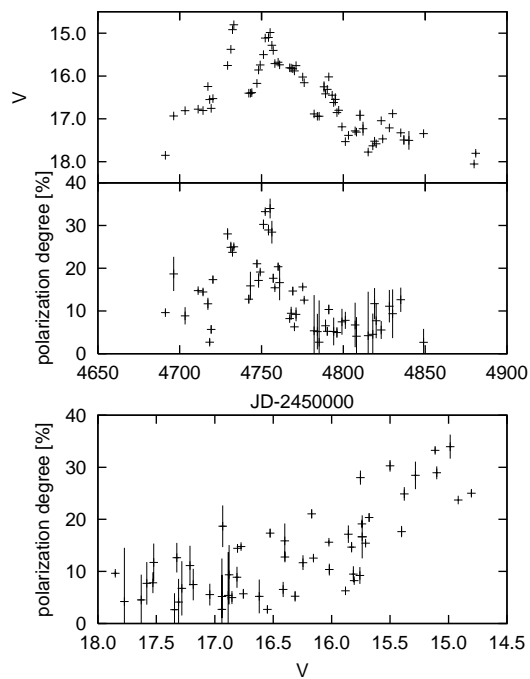}
 \caption{Our results for AO~0235$+$164. Top
 and middle panel: the lightcurve and the polarization degree
 variation. Bottom panel: the relation of the flux and the polarization
 degree.}\label{fig:AO0235}
\end{figure}

\begin{figure}[t]
 \includegraphics[width=130mm]{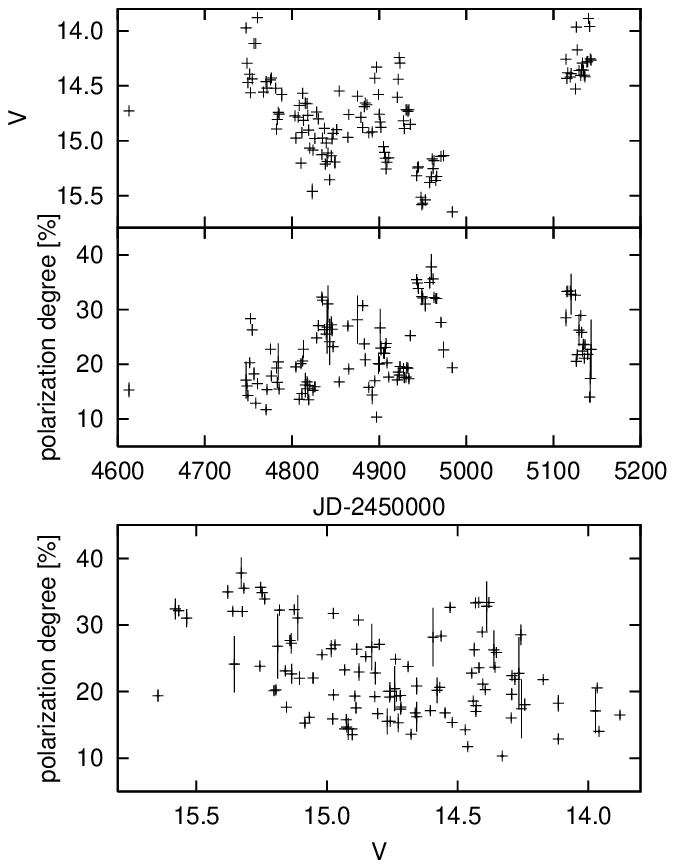}
 \caption{Our results for OJ~287. Top and
 middle panel: the lightcurve and the polarization degree
 variation. Bottom panel: the relation of the flux and the polarization
 degree.}\label{fig:OJ287}
\end{figure}

\begin{figure}[t]
 \includegraphics[width=60mm]{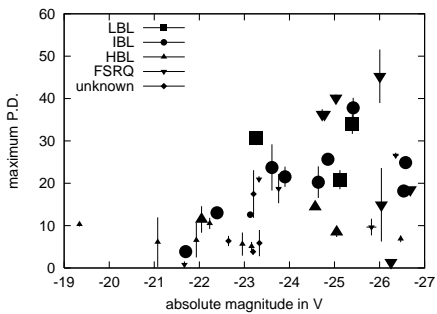}
 \caption{Maximum polarization degree against the $V$-band
 absolute magnitudes of 38 blazars.}\label{fig:abmag_polmax}
\end{figure}

As well as the correlation of the flux and the color, we investigated
 the correlation of the flux and the polarization degree.
We show an example of the light curve and the polarization degree in
 figure~\ref{fig:AO0235}.
This figure shows the results for AO~0235$+$164.
As shown in figure~\ref{fig:AO0235}, the polarization degree of this
 object became high when it was brighter.
We found that 10 blazars exhibited a significant correlation of the
 flux and the polarization degree.

We found that four blazars (QSO~0454$-$234, OJ~49, ON~325, and OJ~287) showed
 a significant anti-correlation of the flux and the polarization degree.
For example, figure~\ref{fig:OJ287} shows the lightcurve and
 the polarization degree of OJ~287.
In this case, the polarization degree was lower when the object was brighter.

As well as in subsection~3.1, the sample reduces to 29 blazars without
 the objects which were observed for $<10$~d.
The numbers of the objects showing significant positive and negative
 correlation correspond to 34~\% and 14~\% of the sample, respectively.
The correlation of the flux is weak in the polarization degree compared
 with that in the color.

Figure~\ref{fig:abmag_polmax} shows the average of the absolute
 magnitudes and the maximum value of the polarization degree of 39 blazars. 
The polarization degree kept low in the blazars with low luminosities,
 like HBLs, while prominent flares of the polarization degree were
 observed in the blazars with high luminosities.
The variation amplitude of the polarization degree was also high/low for
 the high/low-luminosity blazars, since the minimum polarization degree
 reached to 0--5\% even in the high-luminosity objects.

\section{Discussion}
 \subsection{Lightcurve and Color}

\cite{Kirk1998} propose that the bluer-when-brighter trend indicates an
 energy injection event in the emitting region.
As a result of the energy injection, the number of
 high energy electrons increases, and thereby, the synchrotron flux in
 short wavelengths increases more than that in long wavelengths.
The most plausible scenario for the energy injection mechanism is the
 internal shocks between relativistic shells (e.g. \cite{Zhang2002}).
We detected a significant bluer-when-brighter feature in 83~\% objects
 in our sample.
This high fraction suggests that most of variations in blazars are caused by
 the internal shock in the jet.
Since our observation lasted for $\lesssim1$~yr, we have no information about
 the color--flux correlation in variations having a time scale longer
 than $\gtrsim 1$~yr.

A universal relation between the flux and color has not been established
 because several objects were reported to exhibit a redder-when-brighter
 trend or no significant correlation of the flux and color
 (\cite{Gu2006}; \cite{Villata2006}; \cite{Raiteri2008}).
Actually, our observation also detected clear redder-when-brighter
 trends in three objects, while they appeared only in faint states of each
 object.
This redder-when-brighter behavior in the faint states is consistent
 with the scenario that the contribution of the thermal emission from
 the accretion disk becomes strong when the synchrotron emission from the
 jet weakened (\cite{Villata2006}).
Even in our redder-when-brighter objects, our observation revealed that
 the bluer-when-brighter feature appeared when they brightened.
Hence, the bluer-when-brighter trend is presumably a universal
 feature for synchrotron flares in blazars.

An alternative scenario for the variations of the flux is the change in
the beaming factor of the emitting region.
In this scenario, the flux from a beamed emitting region should increase
 in all wavelengths.
Our results show that the variation amplitude of the flux and color
 trend to be smaller in HBL than that in LBL.
This implies that the variation amplitude originated from low-energy
 electrons is small compared with that from high-energy electrons.
Thus, our results support the internal shock scenario rather than the
 scenario with the beaming-factor changes for most variations in blazars.


 \subsection{Lightcurve and Polarization}

The weak correlation of the flux and the polarization degree may be
 caused by the presence of two or more components in the polarization.
Polarization variations correlating with the flux could be hidden in
 the case that there is another polarization component whose temporal
 variation is independent of the total flux.
Based on this picture, we have developed a method to separate the
 observed polarization into two components which correlate and
 not-correlate with the flux variation (\cite{Uemura2010}).
Applying this method to our blazar data, we found that the behavior of
 polarization in several blazars can be explained with this
 two-component picture.

\section{Summary}
We performed photopolarimetric monitoring of 41 blazars in the optical
 and near-infrared bands using the Kanata telescope.
Our findings are summarized below:
 \begin{itemize}
    \item The bluer-when-brighter trend was observed in 24 blazars which
	  correspond to 83~\% of our well-observed sample.
          Our observation strongly indicates that the
	  bluer-when-brighter trend is a universal feature in blazars.
    \item Three blazars showed the redder-when-brighter trend when they
	  were in faint states.
	  Even in those three objects, the bluer-when-brighter trend
	  appeared when they were in bright states.
    \item We found that the variation amplitudes were
	  smaller in the objects with lower luminosities, like HBLs,
	  in the magnitude, color, and polarization degree.
    \item The significant correlation of the flux and the polarization
	  degree was observed in 10 blazars which correspond to 34~\% of
	  our well-observed sample.
	  The correlation with the flux is weak in the polarization
	  compared with that in the color.
\end{itemize}

\bigskip 

\bigskip 

\end{document}